\documentclass[preprint,eqsecnum,aps,showpacs]{revtex4}
\usepackage{dcolumn}
\usepackage{graphicx}
\usepackage{latexsym}
\usepackage{multirow}

\begin{document}

\title{Bulk spacetimes for cosmological braneworlds with a time-dependent 
extra dimension} 

\author{Suman Ghosh\footnote{Electronic address : {\em
suman@cts.iitkgp.ernet.in}}${}^{}$ and Sayan Kar\footnote{Electronic
address : {\em sayan@cts.iitkgp.ernet.in}}${}^{}$}
\affiliation{Department of Physics and Centre for Theoretical Studies
\\ Indian Institute of Technology, Kharagpur 721 302, India}

\begin{abstract}
We explore the possibilities of constructing bulk spacetimes in five dimensions
for warped braneworld models
with a spatially flat Friedmann-Robertson-Walker (FRW) 
line element on the 3-brane and with a
time-dependent extra dimension. Our first step in this direction
involves looking at the status of energy conditions when such
a bulk line element is assumed. We check these conditions
by analysing the relevant inequalities, for specific 
functional forms (chosen to satisfy certain desirable features) 
of the warp factor, the cosmological scale factor and the 
extra-dimensional scale factor.
Subsequently, we aim at obtaining solutions with
different types of bulk matter sources. We begin with a
general analysis of the solution space of non-singular Randall-Sundrum
type bulk models with an exponential warp factor and a chosen equation
of state.  Thereafter, we focus on three specific
bulk sources -- the ordinary scalar field, the
Brans-Dicke scalar and the dilaton.
In each case, we are able to solve the field
equations and obtain desirable solutions for which, we once again
check the viability of the energy conditions. 
We also show how one can place branes in the bulk using the 
junction conditions. The issue of
resolution of the bulk singularities which appear in our
solutions, using standard methods, is also presented briefly.
In summary, we are able to demonstrate, that
it is indeed possible to construct viable bulk spacetimes for 
warped cosmological braneworlds
with a time-varying extra dimension and with bulk matter
satisfying the energy conditions.  
\end{abstract}


\maketitle


\section{Introduction}
Though yet to be seen in experiments, extra dimensions have been
around in theoretical constructs for almost a century now. The fact
that such a seemingly exotic idea has survived for so long seems
to make us feel that there is bound to be some amount of advantage
in having them. The advantage stems from several directions. For instance,
one may ask -- {\em why four dimensions?} -- a question to which one
really does not have an answer. Or, while making models of unification
(such as superstrings \cite{string}) one ends up with a need of more dimensions and the subsequent need for compactification. More recently, we have seen
how the age-old problem --the hierarchy problem--can be solved
(obviously not uniquely) with extra dimensions \cite{rs}.

The most popular among today's models is the one we mentioned at the
end in the last paragraph-- the so-called warped braneworld or Randall-Sundrum
(referred henceforth as RS)
 model. This model assumes a curved spacetime (the bulk) in five dimensions.
The four dimensional piece however, depends on the extra dimension, a feature unique for this class of models (different from the usual
Kaluza--Klein \cite{kk}). Thus, in such a model we have a metric
ansatz:
\begin{equation}
ds^2 = e^{2 f(\sigma)} \left (-dt^2 + d{\vec x}^2\right) + r_c^2 d\sigma^2
\end{equation}
The $\sigma= constant$ (say  $\sigma_0$) section is a four dimensional Minkowski space scaled by
the factor $e^{2 f(\sigma=\sigma_0)}$.

But such a metric ansatz is toy. This is because the metric on
the $\sigma=$ constant surface must necessarily be cosmological, in order
to tally with the universe. Further, there is no reason for us to
assume $r_c$ to be a constant. It could be a function of space and time. Such a spacetime dependence effectively makes the scale of the
extra dimension, a field, commonly known today as the radion. However, here, 
we restrict ourselves to only a temporal dependence of $r_c$. Together, these two modifications, i.e. having
a cosmological on-brane metric and a time dependent extra dimensional scale drive
us towards a more realistic braneworld model.

However, the generalisation of the simple RS line element to the more 
complicated one
(i.e. with a cosmological scale factor and a time dependent extra dimensional factor) makes
life difficult for the general relativist to find meaningful exact solutions. 
Several attempts have
been made so far. Notable among them are discussed in \cite{chung_freese1,
chung_freese, koyama}. Further, the existence of a
time dependent extra dimension in an unwarped scenario has been investigated with the
aim of finding out whether such a time dependence can drive an accelerating universe \cite{unwarped}.
Cosmological applications including implications for the CMB anisotropies \cite{chan_chu} and BBN nucleosynthesis \cite{bbn} have been analysed in recent times. On the other front, warped cosmological
braneworlds with a constant scale for the extra dimension have been studied extensively \cite{statics}
ever since the RS model came into existence.

Our intentions in the investigations here, is to first figure out whether such 
solutions (one can call them generalised brane-bulk systems) exist
and, if they do, how do the matter fields, spacetime geometry behave. 
Our programme is as
follows. In the next section, we write down the field equations. Then we move on to the issue
of energy conditions and try to figure out, if it is possible to have viable 
forms for the unknown metric functions, which satisfy them. Then, we look at the possibilities of solutions
with matter fields. First, we assume a non--singular bulk with an
exponential warp factor and discuss the solution space for a specific
equation of state. Thereafter, we discuss three cases  
with different types of scalar field actions -- ordinary and phantom scalar, Brans--Dicke scalar and
dilaton. We then return to a discussion on the issue of energy conditions 
for our exact solutions. Finally, we 
 discuss the placing of branes using the junction conditions and the 
question of resolution of bulk singularities. 
We conclude the paper with a summary and some general remarks.

\section{The line element ansatz and Einstein's equations}  

A general warped line element in five dimensions is given as:
\begin{equation}
ds^2 = e^{2f(\sigma)} g_{\alpha\beta} dx^\alpha dx^\beta + g_{44} d\sigma^2
\end{equation}
where $g_{\alpha\beta}$ can, in principle be any metric and $g_{44}$ can be a function
of space, time and $\sigma$, not necessarily separable.

The line element we choose to work with is as follows:

\begin{equation}
ds^2 = e^{2 f(\sigma)} \left [-dt^2 +a^2 (t) \left (\frac{dr^2}{1-kr^2} + r^2 d\theta^2 +
r^2 \sin^{2}\theta d\phi^2 \right ) \right ] +\eta^2 (t) d\sigma^2 \label{eq:metric}
\end{equation}

In the above, we have chosen a cosmological line element on the
$\sigma= constant$ hypersurface. $a(t)$ is the cosmological scale factor and $k=0, \pm 1$
correspond to the usual Friedmann--Robertson--Walker (FRW) metric with 
spacelike sections of zero ($R^3$), positive ($S^3$)
and negative ($H^3$) curvatures respectively. As defined before, 
$e^{2 f(\sigma)}$ is the warp factor. The domain of $\sigma$, which is the
extra dimension, could be $-\infty<\sigma<\infty$ though, as we shall see 
later, we may need to consider finite or semi-infinite domains and/or introduce
additional branes in order to avoid inevitable bulk singularities which appear 
in most of our solutions. However, unlike the Randall-Sundrum
model we now have a time-dependent function $\eta(t)$ associated with
the extra dimension. In general, such a function, when it
is dependent on all the four coordinates $(t,r,\theta,\phi)$, is known as
the {\em radion} field. It measures the scale of the extra dimension at 
different spatial and temporal locations in the four dimensional world. In the two-brane RS1 picture, this quantity $\eta$ is known to be the inter-brane distance and must have a stable value in order to make sure that the branes do not collapse on to each other. Note that in the RS model we always had $\eta(t)$ as 
a constant. 

It might be useful at this stage to find out what our expectations are
about the nature of the unknown functions $f(\sigma)$, $a(t)$ and $\eta(t)$.
Guided by the RS solution, we expect $f(\sigma)$ to be such that the
bulk line element is nonsingular w.r.t. the $\sigma$ coordinate, i.e. there are no bulk singularities. This, as we will see, is largely impossible unless we postulate it to be so. On the other hand, the usual singularity in cosmological time t (or,
conformal time) should exist. In other words, it is reasonable to assume a big--bang singularity
in the four dimensional world where $a(t)$ goes to zero at some time
$t$ but at all spatial positions $r$, $\theta$, $\phi$ as well as $\sigma$.
Secondly, we would like to have a solution where, following our
current understanding of cosmology, we have an expanding universe.
This will imply $a(t)$ to be a monotonically increasing function of $t$ which may have a deceleration or
an acceleration (negative or positive second derivatives, respectively). 
On the other hand, we would not like to have a large scale for the 
extra dimension today,
which leads to the choice that $\eta (t)$ should be monotonically
decreasing in time, but never becoming zero at any finite time. This
choice is motivated from old ideas in Kaluza--Klein cosmology where, of course
extra dimensions are always compact. In fact, a growing extra dimension is 
still fine if it stabilises to a certain value at later times. But, as we will see, the solutions found are essentially of power 
law type which, if growing, will keep growing forever. Thus, we
choose the extra dimension scale to be of the decaying type. 
We mention that we never use the above choices as constraints while solving 
the Einstein equations. The above choices merely imply that we {\em prefer} 
to have an expanding four dimensional universe with a non-compact but shrinking (in time) extra dimensional scale. 
 
One may argue, following the example in \cite{koyama},
that an increasing $\eta(t)$ is an acceptable model 
essentially because the bulk solution can be transformed to a static, conformally flat metric 
by a suitable coordinate transformation (see \cite{koyama}). In such a
case, a growing extra dimensional scale is permissible because it 
corresponds to the motion of the brane (through its embedding) in a static
bulk. 
However, a coordinate transformation  of the type used in \cite{koyama}
seems possible only if the Weyl tensor for the bulk geometry vanishes. 
One can check that, for the metric ansatz \ref{eq:metric} (with $k = 0$), 
all the components of the Weyl tensor vanish if the following constraint 
is satisfied,
\begin{equation}
\frac{\ddot a}{a} - \frac{\dot a^2}{a^2} + \frac{\dot a \dot\eta}{a\eta} - \frac{\ddot\eta}{\eta} = 0. \label{eq:weyl_constraint}
\end{equation}
For solutions of power law type, we may choose, $a(t) \sim t^{\nu_1}$ 
and $\eta(t) \sim t^{\nu_2}$. Eq. \ref{eq:weyl_constraint} then implies
\begin{equation}
(\nu_1 - \nu_2) (\nu_2 - 1) = 0. \label{eq:weyl_constraint1}
\end{equation}
$\nu_1 = \nu_2$ means a trivial solution which results in the 
same functional form for both the scale factors and $\nu_2 = 1$ is 
essentially what was found in \cite{koyama}. 
It may be noted that, for the scale factors obtained as solutions in
our examples, the above constraint relation does not hold and thus, the
bulk geometries are not conformally flat. 

Furthermore, following the
discussion in \cite{chung_freese1}, a moving brane arises when we
make a gauge choice in which the
metric on the $\sigma-t$ section is manifestly conformally flat. 
Fixing the brane implies that we do not retain the full gauge freedom. 
Thus, the growing/decaying nature of the extra dimension is linked
with the motion of the brane in the bulk. By making a choice of
decaying extra dimensions, we therefore, do not eliminate the
possibility of viable models with growing extra dimensions though
our preference (and the obtained solutions) is based on some logic-- 
essentially linked with
old Kaluza--Klein ideas and the issue of a stabilised extra
dimensional scale at later times. We mention that such solutions with a 
decaying extra dimensional scale have been discussed earlier, for example in
\cite{chung_freese1} itself.  

Thus, our central question now is: do the Einstein field equations with
a matter source have such kind of solution that we seek? To answer this query, we must
write down these equations. We first choose to write down the Einstein tensor and perform our analysis by
assuming it to be the {\em required} stress energy for specific choices of 
the unknown functions.
This, in the older literature of general relativity is known as Synge's {\it g-method} \cite{synge}. The
other way, is to actually solve the Einstein's equations with some specific choices of matter
stress energy (eg. different types of scalar fields residing in the bulk) -- this, following Synge
is the {\it T-method}.

The nonzero components of Einstein tensor (in the frame basis) 
for the metric \ref{eq:metric} with $k = 0$ (i.e. a spatially flat cosmological brane) are,
\begin{eqnarray}
G_{00} &=&  e^{-2f}\left (3\frac{\dot a^2}{a^2} + 3 \frac{\dot a}{a} \frac{\dot \eta}{\eta} \right ) - \frac{1}{\eta^2} \left ( 3 f'' + 6 f'^2 \right ) \\
G_{\alpha \alpha} &=& - e^{-2f} \left (2 \frac{\ddot a}{a} + \frac{\dot a^2}{a^2} + 2 \frac{\dot a}{a} \frac{\dot \eta}{\eta} + \frac{\ddot \eta}{\eta}  \right ) +\frac{1}{\eta^2} \left ( 3 f'' + 6 {f'}^2 \right ) \\
G_{44} &=& - e^{-2f} \left (3 \frac{\ddot a}{a} + 3 \frac{\dot a^2}{a^2}\right ) +\frac{6 f'^2}{\eta^2} \\
\mbox{and} \hspace{1cm}G_{04} &=& 3\frac{\dot \eta}{\eta^2} f' e^{-f} 
\end{eqnarray}
where, in the above, $G_{\alpha\alpha}$ refers $G_{11}, G_{22}, G_{33}$, the
three spatial indices in the 4D FRW metric. A dot denotes differentiation
w.r.t time while a prime indicates differentiation w.r.t. $\sigma$.

Defining new variables, $\frac{\dot a}{a} = x$, $\frac{\dot \eta}{\eta} = y$, 
we have,
\begin{eqnarray}
G_{00}  &=& e^{-2f} ( 3 x^2 + 3 x y ) - \frac{1}{\eta^2}( 3 f''+ 6 f'^2 ) \nonumber  \\
G_{\alpha\alpha} &=& -e^{-2f}(2\dot x + 3x^2 + 2 x y + \dot y + y^2) + \frac{1}{\eta^2}( 3 f'' + 6 f'^2) \nonumber  \\
G_{44} &=&  - e^{-2f}(3 \dot x + 6 x^2) + \frac{1}{\eta^2}{6f'^2}  \nonumber  \\
\mbox{and} \hspace{1cm} G_{04} &=& 3 y f' \frac{e^{-f}}{\eta} \label{eq:ETcomps}
\end{eqnarray}

Some interesting features in the structure of these
equations may be noted here.

\begin{itemize}
\item{The presence of the flux term emerging out of $G_{04}$ which is 
zero only when $y=\frac{\dot \eta}{\eta}$ is zero and/or when
the geometry is unwarped (i.e. $f$ is a constant).}
\item{The separability of functions of time and $\sigma$
 which enables us to make an attempt towards 
constructing solutions by equating coefficients of $e^{-2f}$ and $\frac{1}{\eta^2}$ on both
sides of the Einstein's equations with some matter source.}
\item{A vacuum solution ($G_{ij}=0$) turns out to be either un-warped or with a
time independent extra dimension. Also, with a bulk cosmological 
constant, (i.e. $G_{ij}+\Lambda g_{ij}=0$), a non-trivial solution cannot be 
found.}
\end{itemize}

\section{The status of energy conditions}

Given the above expressions for the Einstein tensors, we can now ask
the question: {\em is it possible to have geometries, i.e. specific functional forms of $a(t)$, $\eta(t)$ and $f(\sigma)$, with matter satisfying the weak or null energy conditions?} To do this analysis we need to know about the energy conditions in a bit more detail, largely because certain non–trivialities arise because of a non–zero $G_{04}$ term (equivalently, a non–zero $T_{04}$ must also be there) being present.

\subsection{Energy conditions: weak and null}
 
The Weak Energy Condition is given by,
\begin{equation}
G_{IJ}U^IU^J \ge 0 \label{eq:WEC}
\end{equation}
where $U^I$ is a nonspacelike vector and a similar inequality defines 
then Null Energy Condition if $U^I$ is a null vector.

One can write down the energy conditions in terms of the eigenvalues of the energy-momentum tensor (which must be real). If $\lambda_0$ denotes the eigenvalue corresponding to the timelike eigenvector, the weak energy condition is equivalent to the following simple relations amongst the eigenvalues \cite{wald}.
\begin{equation}
- \lambda_0 \ge 0 \hspace{.5cm}\mbox{and}\hspace{.5cm}  - \lambda_0 +\lambda_\alpha \ge 0  \hspace{.5cm}\mbox{where}\hspace{.5cm} (\alpha = 1,2,3,4)  . \label{eq:WECsimplified}
\end{equation} 
In effect, we find the inequalities by diagonalising the energy-momentum tensor.
The eigenvalues of the energy-momentum tensor are the roots of the equation
\begin{equation}
|G_{IJ} - \lambda g_{IJ}| = 0 \label{eq:chacteristic}
\end{equation}
which reduces to, 
\begin{equation}
\left| \begin{array}{ccccc}
\rho + \lambda  &  0  &  0  &  0  &  q \\
0     & p_1 - \lambda &  0  &  0  &  0 \\
0     &  0  & p_1 - \lambda &  0  &  0 \\
0     &  0  &  0  & p_ 1- \lambda &  0 \\
q     &  0  &  0  &  0  & p_2 - \lambda
\end{array} \right| = 0 \label{eq:eigen}
\end{equation}
where,
\begin{eqnarray}
\rho =  - \frac{3 f'' + 6 f'^2 }{\eta^2} + e^{-2f} \left(3\frac{\dot a^2}{a^2} + 3 \frac{\dot a}{a} \frac{\dot \eta}{\eta} \right), 
\end{eqnarray}
\begin{eqnarray}
p_1 = \frac{1}{\eta^2} \left ( 3 f'' + 6 f'^2 \right ) -  e^{-2f} \left(2 \frac{\ddot a}{a} + \frac{\dot a^2}{a^2} + 2 \frac{\dot a}{a} \frac{\dot \eta}{\eta} + \frac{\ddot \eta}{\eta}  \right),
\end{eqnarray}
\begin{equation}
p_2 = \frac{ 6 f'^2}{\eta^2} - e^{-2f} \left (3 \frac{\ddot a}{a} + 3 \frac{\dot a^2}{a^2}\right )
\end{equation}
\begin{equation}
\mbox{and}\hspace{.5cm} q  = 3\frac{\dot \eta}{\eta^2} f' e^{-2f}.
\end{equation}
Here, and henceforth, we shall absorb the factor $\kappa$ in the
Einstein equation $G_{IJ} = \kappa T_{IJ}$ through appropriate
scaling redefinitions. The five eigenvalues from Eq. \ref{eq:eigen} assume the following forms,
\begin{equation}
\lambda_0 = \frac{1}{2} (- \rho + p_2 - \sqrt{(\rho + p_2)^2 - 4q^2}) 
\end{equation}
\begin{equation}
\lambda_1 = \lambda_2 = \lambda_3 = p_1 
\end{equation}
\begin{equation}
\lambda_4 = \frac{1}{2} (- \rho + p_2 + \sqrt{(\rho + p_2)^2 - 4q^2}) 
\end{equation}
The fact that the above eigenvalues have to be real, leads to the following requirment
\begin{equation}
(\rho + p_2)^2 - 4q^2 \ge 0 \label{eq:bounds}. 
\end{equation}
This may restrict the allowed domains of $t$ and $\sigma$. 
As the eigenvector corresponding for the eigenvalue $\lambda_0$ is timelike 
or null, the effective inequalities for Weak Energy Condition turns out to be
\cite{santos}, 
\begin{equation}
F_1 = \rho - p_2 + \sqrt{(\rho + p_2)^2 - 4q^2} \ge 0 , \label{eq:ineq1}
\end{equation} 
\begin{equation}
F_2 = \rho + 2 p_1 - p_2 + \sqrt{(\rho + p_2)^2 - 4q^2} \ge 0 , \label{eq:ineq2}
\end{equation} 
\begin{equation}
\mbox{and}\hspace{.5cm} F_3 = \sqrt{(\rho + p_2)^2 - 4q^2} \ge 0 \label{eq:ineq3}.
\end{equation} 

\begin{figure}[!ht]
\includegraphics[width = 5 in]{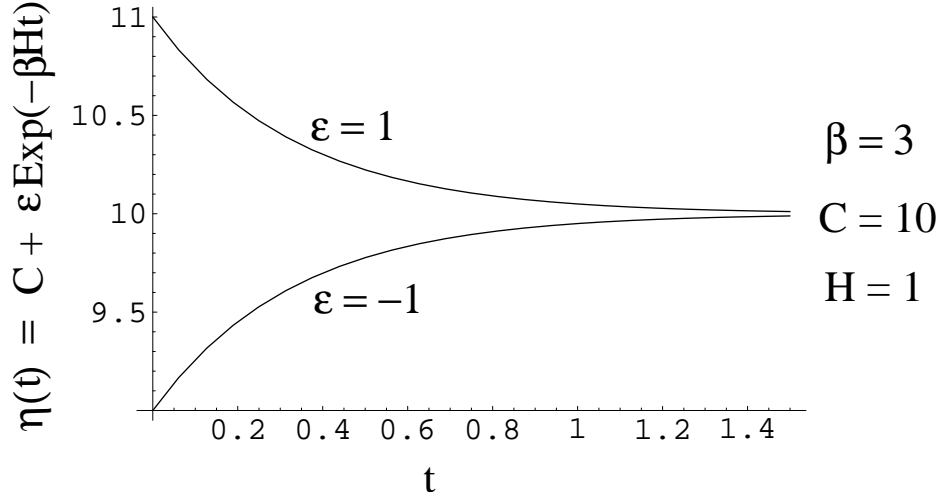}
\caption{variation of $\eta(t)$ for $\epsilon = 1$ and $\epsilon = -1$.} \label{fig:eta} 
\end{figure}

\begin{figure}[!ht]
\includegraphics[width = 5 in]{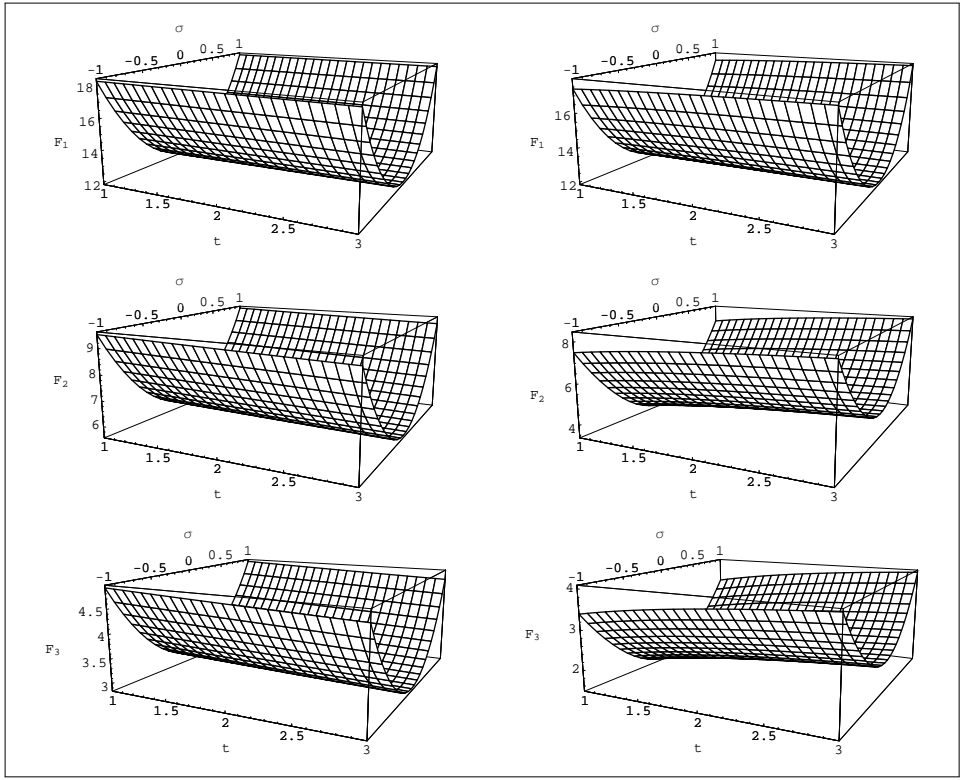}
\caption{Status of the inequalities for a de-Sitter brane for following set of parameter values: $b = \frac{1}{2}$, $c = 2$, $\beta = 3$, $H = 1$, $\epsilon = 1$ (left column) and $\epsilon = -1$ (right column).} \label{fig:plot1} 
\end{figure}

\begin{figure}[!ht]
\includegraphics[width = 5in]{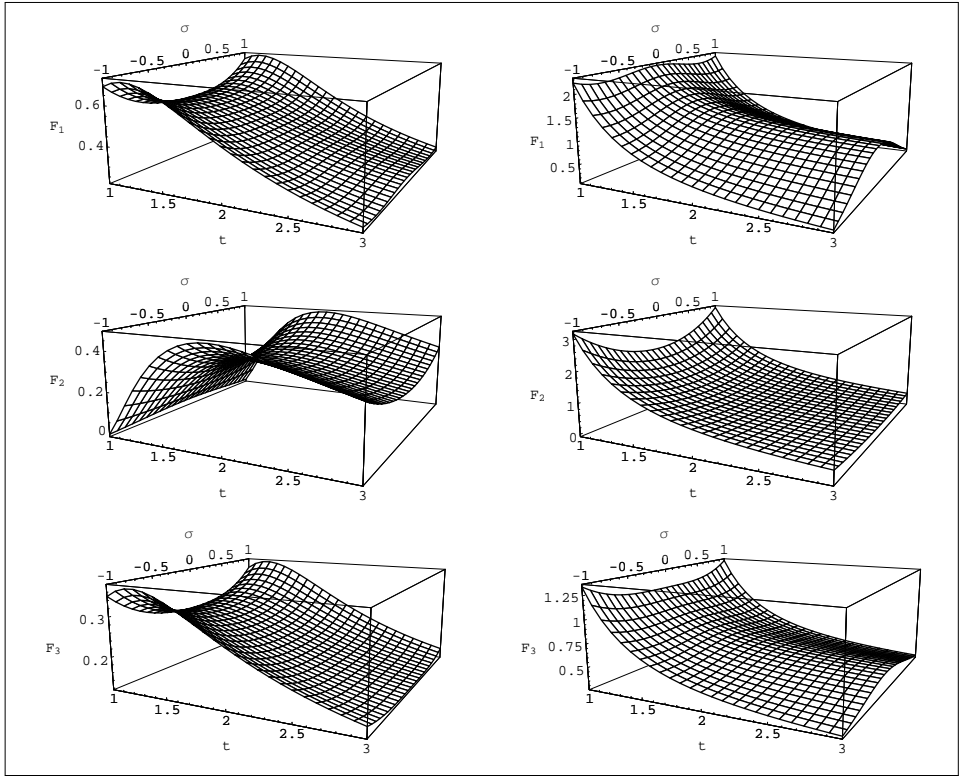}
\caption{Status of the inequalities for a radiative brane for following set of parameter values: $b = \frac{1}{2}$, $c = 6$, $\beta = 2.4$, $H = 1$, $t_0 = 1$, $\epsilon = 1$ (left column) and $\epsilon = -1$ (right column).} \label{fig:plot2} 
\end{figure}

Fig.\ref{fig:plot1} shows status of the above three inequalities for two models in the two columns,
\begin{eqnarray}
\mbox{where}\hspace{1cm} f(\sigma) &=& - b\log (cosh\ \sigma) \hspace{1cm}\mbox{represents a thick brane},\nonumber\\
a(t) &\sim& e^{Ht} \hspace{1cm}\mbox{a de-Sitter brane} \label{eq:test1} \\
\mbox{and} \hspace{1cm} \eta(t) &=& c + \epsilon e^{-\beta H t} \nonumber
\end{eqnarray}
(where $t$ and $\sigma$ run from $0$ to $\infty$ and $-\infty$ to $\infty$ respectively) 
with $\epsilon = 1$ and $\epsilon = -1$ respectively (variation of $\eta(t)$ with two different values of $\epsilon$ is shown in Fig.\ref{fig:eta}). Fig.\ref{fig:plot2} shows the variations of the same functions (i.e. $F_{1,2,3}$) for 
\begin{eqnarray}
a(t) &\sim& \sqrt{t}  \hspace{1cm}\mbox{a radiative brane}\nonumber\\
\mbox{and} \hspace{1cm}\eta(t) &=& c + \epsilon e^{-\beta H(t-t_0)} \label{eq:test2} 
\end{eqnarray}
where $t$ runs from $t_0$ ($=1$ say) to $\infty$ and the warp factor is the same as in Eq. \ref{eq:test1}.

We find that for the above two models, all the inequalities are satisfied at 
least with the specific choices of parameters we have made. It has been 
checked that these inequalities can be satisfied in the presence of a 
decaying warp factor. However, for a growing warp factor (i.e. for a negative $b$) the inequalities are violated at least in some spacetime region.

\section{Exact analytic solutions in the bulk}
In this section, we discuss possible exact solutions in the bulk. We first 
assume an exponential warp factor and discuss some specific consequences
for a given equation of state.
In subsequent sub--sections, we look at exact solutions with different
types of scalar fields -- an ordinary massless scalar, the Brans-Dicke scalar and finally a dilatonic
scalar. 

\subsection{Models with an exponential (Randall--Sundrum type) warp factor}

In the Randall-Sundrum scenario, the warp factor is of the form
$f(\sigma)=-b \vert \sigma \vert$. Using the functional form
$f(\sigma)= -b \sigma$ (or $b\sigma$) and restricting ourselves to the 
$\sigma \geq 0$
(or $\sigma \leq 0 $) region of the extra dimension, we now investigate
some special cases by looking at the Einstein tensors and the resulting
required matter stress-energy. To make things
more quantitative, let us first write down the Einstein tensors with
the above-mentioned warp factor.
\begin{eqnarray}
G_{00}  &=&  e^{2 b \sigma} ( 3 x^2 + 3 x y ) - \frac{6 b^2}{\eta^2},\nonumber \\
G_{\alpha\alpha} &=& - e^{2 b \sigma}(2\dot x + 3x^2 + 2 x y + \dot y + y^2) + \frac{6b^2}{\eta^2}, \nonumber \\
G_{44} &=& - e^{2 b\sigma}(3 \dot x + 6 x^2) + \frac{6 b^2}{\eta^2} \nonumber \\
\mbox{and} \hspace{1cm} G_{04} &=& 3 y b \frac{e^{b \sigma}}{\eta}
\end{eqnarray}

\begin{figure}[!ht]
\includegraphics[width = 3in]{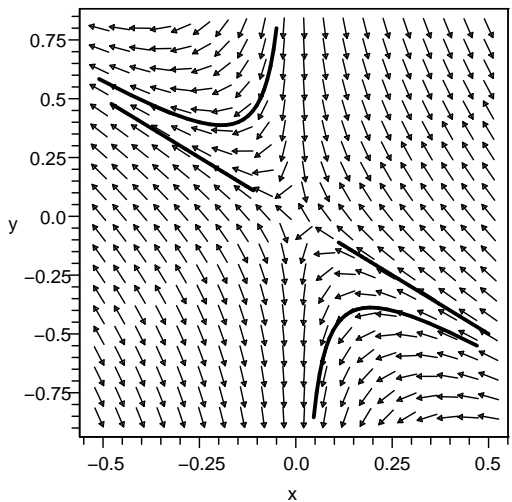}
\caption{The solution space for the dynamical system in Eq. 4.3. 
The solid lines indicate different flows in the solution space of \{$x$,$y$\} for the following different initial conditions: $\{x(0),y(0)\} = (0.5,-0.5), (0.47,-0.55), (-0.05,0.8)$ and $(-0.11,0.11)$, for $t$ running from $0$ to $3.5$.} \label{fig:solspace} 
\end{figure}

\noindent (a) Let us first look at the situation where we have
\begin{equation}
G_{00}=-G_{\alpha\alpha}=-G_{44}.
\end{equation}
Since the above relation holds for the factors associated with the $\frac{1}{\eta^2}$ terms in the Einstein tensors, conditions on $x$, $y$ and its derivatives
emerge when we impose identical conditions for the factors associated
with the $e^{2b \sigma}$ terms.
These conditions, after some elementary algebra lead to:
\begin{eqnarray}
\dot x &=& -x^2 + x y, \nonumber \\
\dot y &=& 2 x^2-x y - y^2. \label{eq:autonomus1}
\end{eqnarray} 
Eq. \ref{eq:autonomus1} is an autonomous dynamical system of nonlinear first 
order differential equations. The general feature of all the solutions can be 
determined through a solution space analysis \cite{strogatz} of the above 
system. In this case, the line $x = y$ is a critical curve, i.e. every point
on this line (except the origin) is a non-isolated fixed point. The eigenvalues of the Jacobian matrix (of the linearised system) at any such fixed point 
($x*,y*$) are ($0, -4x*$), which suggests, for $x*>0$ fixed points are 
neutrally stable whereas for $x*<0$ they are neutrally unstable. It is 
difficult to make any conclusive comments, analytically, on the behaviour 
of phase space trajectories near those points, because such borderline cases 
are very sensitive to nonlinear terms in the equations. But the phase portrait 
(Fig.\ref{fig:solspace}) indeed confirms our conclusions about the nature of 
the fixed points. Note that, the ($0,0$) point in the phase space is irrelevant as the Jacobian itself becomes singular and therefore it does 
not fall in any category of fixed points. From a physical point of view, 
the origin represents a static universe. We will see that no trajectory reaches that state in finite time. This provides a justification for the origin not 
representing a viable solution. Now, we turn our attention to 
the physically most preferable region of the solution space. This is the 
below--right quadrant, where $x$ is positive and $y$ is negative. There is 
one such curve, $x = -y$, which is distinct from every other. It approaches 
the origin ($0,0$), i.e. towards a static universe, as $t\rightarrow \infty$.
Any  $(x,y)$ combination that is located above that line flows toward the line $x=y$, resulting in the same scale factor for both kind of spatial dimensions, at
 $t = \infty$. On the other hand, flows located below the curve $x = -y$, 
tend toward a static brane ($x = 0$) with an extra dimension whose size 
shrinks, with ever increasing rate ($y \rightarrow -\infty$), towards zero 
as $t\rightarrow \infty$.

Following the phase portrait, we assume $y=n x$, which implies, for consistency, $n^2=1$ or $n=\pm 1$.
The n=1 case is trivial because it leads to constant $a (t)$ and hence,
constant $\eta(t)$. With $n=-1$ we obtain, $a(t)\sim t^{\frac{1}{2}}$ and
$\eta(t) \sim t^{-\frac{1}{2}}$. This specific solution represents a radiative brane while the extra dimesion is indeed decaying (in accordance with our preference). For these forms of the $a(t)$ and $\eta(t)$
the stress energy is remarkably simple because, 
the coefficients of $e^{2b \sigma}$ in the Einstein tensors are all
zero. Therefore, we just have:
\begin{equation}
T_{00}=-\frac{6b^2}{t}=-T_{\alpha\alpha}=T_{44}, T_{04}=\frac{3 b}{t}.
\end{equation}

\noindent (b) We now move on to another case where we choose $a(t)=e^{\beta t}$
and $\eta (t) = e^{-\gamma t}$ with $\beta, \gamma >0$ while $f(\sigma)$ remains
the same (i.e. $f(\sigma)\propto \sigma$). The Einstein tensors now take the form 
\begin{eqnarray}
G_{00}  &=&  e^{2b \sigma} ( 3 \beta^2 - 3 \beta \gamma ) - \frac{6 b^2}{e^{-2\gamma t}}, \nonumber\\
G_{\alpha\alpha} &=& - e^{2 b \sigma}( 3\beta^2 - 2 \beta \gamma  + \gamma^2) + \frac{6b^2}{e^{-2\gamma t}}, \nonumber \\
G_{44} &=& - e^{2 b \sigma}( 6 \beta^2) + \frac{6 b^2}{e^{-2\gamma t}},  \nonumber \\
G_{04} &=& -3 \gamma b \frac{e^{b \sigma}}{e^{-\gamma t}}.
\end{eqnarray}
Notice that if $\beta=-\gamma$, we end up with the same relations between
the Einstein tensors as in (a) above, though this involves growing 
factors for both the cosmological scale and the extra dimension scale. 
In addition, we also note that with $\gamma=0$ one obtains a de-Sitter
(or anti de-Sitter) brane with a constant scale for the extra dimension. 

We now move on towards obtaining solutions in the true sense by assuming
specific forms of the bulk energy momentum tensor.

\subsection{Bulk ordinary scalar}
The simplest choice for a bulk energy momentum tensor is that of an
ordinary, massless scalar field, for which we have:
\begin{equation}
T_{IJ}^{scalar} = \partial_I \phi \partial_J \phi - \frac{1}{2}g_{IJ} \partial_K\phi \partial^K \phi  \label{eq:Tijord}
\end{equation}
with components, such as
\begin{eqnarray}
T_{00}  = \frac{e^{-2f}}{2} \dot\phi^2 + \frac{\phi'^2}{2\eta^2} = T_{44}, \hspace{2cm} \nonumber \\
T_{\alpha\alpha}  = \frac{e^{-2f}}{2} \dot\phi^2 - \frac{\phi'^2}{2\eta^2} \hspace{.5cm} \mbox{and} \hspace{.5cm}
T_{04} =  \dot\phi \phi' \frac{e^{-f}}{\eta}. \label{eq:TijOrd} 
\end{eqnarray}
We now need to equate the above with the Einstein tensors and obtain
solutions.
To keep things simple, let us assume
\begin{equation}
\phi(t,\sigma) \equiv \phi_1(t) + \phi_2(\sigma). \label{eq:phiformord} 
\end{equation}
From the equation for the $04$ component, we note that $\dot\phi_1 = m y$ and $\phi_2' = \frac{3}{m} f'$. So, the form of $\phi (t, \sigma)$ will be
\begin{equation}
\phi(t,\sigma) = m\log[\eta(t)] + \frac{3}{m} f(\sigma) + const.  \label{eq:phisolord} 
\end{equation}
Using the relations between the coefficients of $\frac{1}{\eta^2}$ on both sides gives:
\begin{equation}
f(\sigma) =\frac{1}{4} \log (\sigma-\sigma_0) + constant \hspace{0.2in};
\hspace{0.2in} {f'}^2 = 12 {\phi_2^{'}}^2.
\end{equation}
In the same way, using the relations between the coefficients of $e^{-2f}$ one gets:
\begin{eqnarray}
\dot x &=& -3 x^2 - x y, \nonumber \\
\dot y &=& -y^2- 3 x y.    \label{eq:autonomus2}
\end{eqnarray}
We now further assume $y= n x$ which is consistent, for all n, with the
above two equations for $\dot x$ and $\dot y$. Therefore, solving
the above two equations one finds
\begin{equation}
a(t) \sim (t - t_1)^{\frac{1}{n+3}} \hspace{0.2in} ; \hspace{0.2in} \eta (t)\sim (t - t_2)^{\frac{n}{n+3}},
\end{equation}
where $t_1$ and $t_2$ are integration constants. However, overall consistency requires that $m^2= \frac{3}{4}$ and 
$n=4\pm 2\sqrt{6}$. For $n=4+2\sqrt{6}$, both $a(t)$ and $\eta(t)$ are 
growing functions of time, which is a feature not desirable. On the other hand, for $n=4-2\sqrt{6}$,
we have the proper behaviour for $a(t)$ and $\eta(t)$. The values of the
exponents for $a(t)$ and $\eta(t)$ are quoted in Table 1. It is worth 
mentioning here that solutions of Eq. \ref{eq:autonomus2} are constrained by 
a consistency requirement of type $x \propto y$, with a specific proportionality constant, which leaves us with solutions very few in number with respect 
to what we obtained in the case discussed earlier. Thus, a solution space 
analysis of Eq. \ref{eq:autonomus2} becomes irrelevant in this case. In fact, 
similar conclusions apply to the other two cases analysed below.  

The case of the phantom scalar (with a negative kinetic energy) has also
been investigated. Using the same methods, we find that solutions do not
exist, primarily because, we end up having ${\phi'}^2=-12 \beta^2 f'^2$, which is
impossible, unless both sides are identically zero.

\subsection{Bulk Brans--Dicke scalar}

Brans--Dicke theory \cite{bransdicke} is well--known as an alternative
theory of gravity where a scalar field $\phi$ is assumed to be
responsible for generating the gravitational constant $G$. Though
experimentally almost ruled out, it serves as a useful model and,
as mentioned later (see next section on dilaton gravity), it has also reappeared
in various contexts in recent times. There have also been a few attempts \cite{mikhailov,majumdar} towards constructing warped braneworld models in 
five-dimensional Brans-Dicke theory.

The action for Brans-Dicke gravity (in five dimensions and without a potential)
where the Brans--Dicke scalar and
the metric are the basic fields, is given as,
\begin{equation}
S = \frac{1}{16 \pi} \int d^5x \sqrt{-g} \left[\phi R - \frac{\omega}{\phi} \phi_{,K}\phi_,^K  +{\cal L}_{matter}\right ] \label{eq:actionbd}
\end{equation}
In the above, we have assumed additional matter other than the
scalar field itself (which, in a sense is not really matter, as such).
The resulting Einstein's equations and the scalar field equation are given as,
\begin{equation}
G_{IJ} = \frac{8\pi}{\phi} T_{IJ} + \frac{\omega}{\phi^2}\left[\phi_{,I}\phi_{,J} - \frac{1}{2} g_{IJ}\phi_{,K}\phi_,^K\right] + \frac{\phi_{;I;J} - g_{IJ}\Box\phi}{\phi}, \label{eq:einsteinbd}
\end{equation}
\begin{equation}
2\omega \frac{\Box \phi}{\phi} + R - \omega \frac{(\nabla\phi)^2}{\phi^2} = 0. \label{eq:scalareqnbd1}
\end{equation}
Contracting Eq. \ref{eq:einsteinbd} and substituting in Eq. \ref{eq:scalareqnbd1}, we get
\begin{equation}
\Box \phi = \frac{8\pi}{4 +3 \omega} T \label{eq:scalareqnbd2}
\end{equation}
where $T$ denotes the trace of the matter energy momentum tensor.

\subsubsection {Solutions with perfect fluid bulk matter}

Let us choose the bulk matter energy momentum tensor to be that of a perfect fluid, $T_{IJ} = diag(\rho, p_{\alpha}, p_{\alpha}, p_{\alpha}, p_{\sigma})$, with a vanishing trace
\begin{equation}
T = - \rho + 3 p_{\alpha} + p_{\sigma} = 0 .\label{eq:tracecond}
\end{equation}
 Thus the scalar field equation essentially becomes
\begin{equation}
\Box \phi = - e^{-2f}\left[\ddot\phi + \left(3\frac{\dot a}{a} + \frac{\dot\eta}{\eta}\right)\dot\phi\right] + \frac{1}{\eta ^2}\left[\phi'' + 4f'\phi'\right] = 0 \label{eq:scalareqnbd_vac}
\end{equation}

Then assuming $\phi(t,\sigma) = \phi_1(t)\phi_2(\sigma)$, the Einstein's equations yield,
\begin{eqnarray}
G_{00} = e^{-2f} \left(\frac{\ddot\phi_1}{\phi_1} + \frac{\omega}{2}\frac{\dot\phi_1^2}{\phi_1^2}\right) - \frac{1}{\eta^2}\left(-\frac{\omega}{2}\frac{\phi_2'^2}{\phi_2^2} + f'\frac{\phi_2'}{\phi_2}\right) + 8\pi \frac{\rho}{\phi},\nonumber
\end{eqnarray}
\begin{eqnarray}
G_{\alpha\alpha} = e^{-2f} \left(\frac{\omega}{2}\frac{\dot\phi_1^2}{\phi_1^2} - x \frac{\dot\phi_1}{\phi_1}\right) +  \frac{1}{\eta^2}\left(-\frac{\omega}{2}\frac{\phi_2'^2}{\phi_2^2} + f'\frac{\phi_2'}{\phi_2}\right)  + 8 \pi \frac{p_{\alpha}}{\phi},  \nonumber
\end{eqnarray}
\begin{eqnarray}
G_{44}  = e^{-2f}\left(\frac{\omega}{2}\frac{\dot\phi_1^2}{\phi_1^2} - y \frac{\dot\phi_1}{\phi_1}\right)  + \frac{1}{\eta^2}\left(\frac{\phi_2''}{\phi_2} + \frac{\omega}{2}\frac{\phi_2'^2}{\phi_2^2}\right)   + 8 \pi \frac{p_{\sigma}}{\phi}, \nonumber
\end{eqnarray}
\begin{equation}
\mbox{and} \hspace{1cm} G_{04} = \frac{e^{-f}}{\eta}\left[(\omega + 1) \frac{\dot\phi_1 \phi_2'}{\phi_1 \phi_2} - f'\frac{\dot\phi_1}{\phi_1} - y \frac{\phi_2'}{\phi_2}\right]. \label{eq:TijBD}
\end{equation}

Using the expressions for the Einstein tensor components, as given in Eq. \ref{eq:ETcomps}, the off-diagonal term in the Einstein equations leads to the following possibility,
\begin{equation}
\frac{\dot\phi_1}{\phi_1} = m y \hspace{.5cm}\mbox{and}\hspace{.5cm} \frac{\phi_2'}{\phi_2} = n f' \hspace{.5cm}\mbox{where}\hspace{.5cm} m = \frac{3 + n}{\frac{4n}{3}-1}. \label{eq:phiconstraintBD}
\end{equation}
Thus, we have,
\begin{equation}
\phi(t,\sigma) \sim \eta^m(t) e^{n f(\sigma)}.  \label{eq:phisolBD}
\end{equation}

Now, dividing Eq. \ref{eq:scalareqnbd_vac} by $\phi$, 
separately equating the coefficients of $e^{-2f}$ and $\frac{1}{\eta^2}$ 
to zero and using \ref{eq:phiconstraintBD} we get,
\begin{eqnarray}
\dot y + 3 x y + (m+1) y^2 &=& 0  \label{eq:scalarBDtime}\\
\mbox{and}  \hspace{1cm} f'' + (n+4) f'^2 &=& 0  \label{eq:scalarBDsigma}
\end{eqnarray}

Similarly, using diagonal components of Einstein equations 
and Eq. \ref{eq:scalarBDtime}, tracelessness of $T_{IJ}$ (Eq. \ref{eq:tracecond}) leads to,
\begin{eqnarray}
\dot x + 2 x^2 - A y^2 = 0, \hspace{1cm}\mbox{where}  \hspace{1cm} A = \frac{m}{3}\left(1 - \frac{m\omega}{2}\right)\label{eq:traceBDtime}\\
\mbox{and}  \hspace{1cm} (12 - n) f'' + \left(\frac{3\omega}{2}n^2 - n^2 - 4 n + 30\right) f'^2 = 0  \label{eq:traceBDsigma}
\end{eqnarray}
Consistency requirement between Eq. \ref{eq:scalarBDsigma} and Eq. \ref{eq:traceBDsigma} gives rise to the following condition,
\begin{equation}
n_{\pm} = \frac{4 \pm 2\sqrt{4 + 3\omega}}{\omega}   \hspace{1cm}\mbox{which also implies} \hspace{1cm} m_{\pm} = \frac{3}{\pm \frac{10}{\sqrt{4 + 3\omega}} - 1}.  \label{eq:nm}
\end{equation}
Then Eq.  \ref{eq:scalarBDsigma} results in
\begin{equation}
f(\sigma) = const. + \frac{1}{4 + n} \log(\sigma - \sigma_0).
\end{equation}
It may be noted that, the factor $\frac{1}{4 + n}$ is always positive for $n = n_-$, but it can be negative as well for $n = n_+$, which implies both 
growing and decaying warp factor solutions are possible.

\begin{figure}[!ht]
\includegraphics[width = 2.25in]{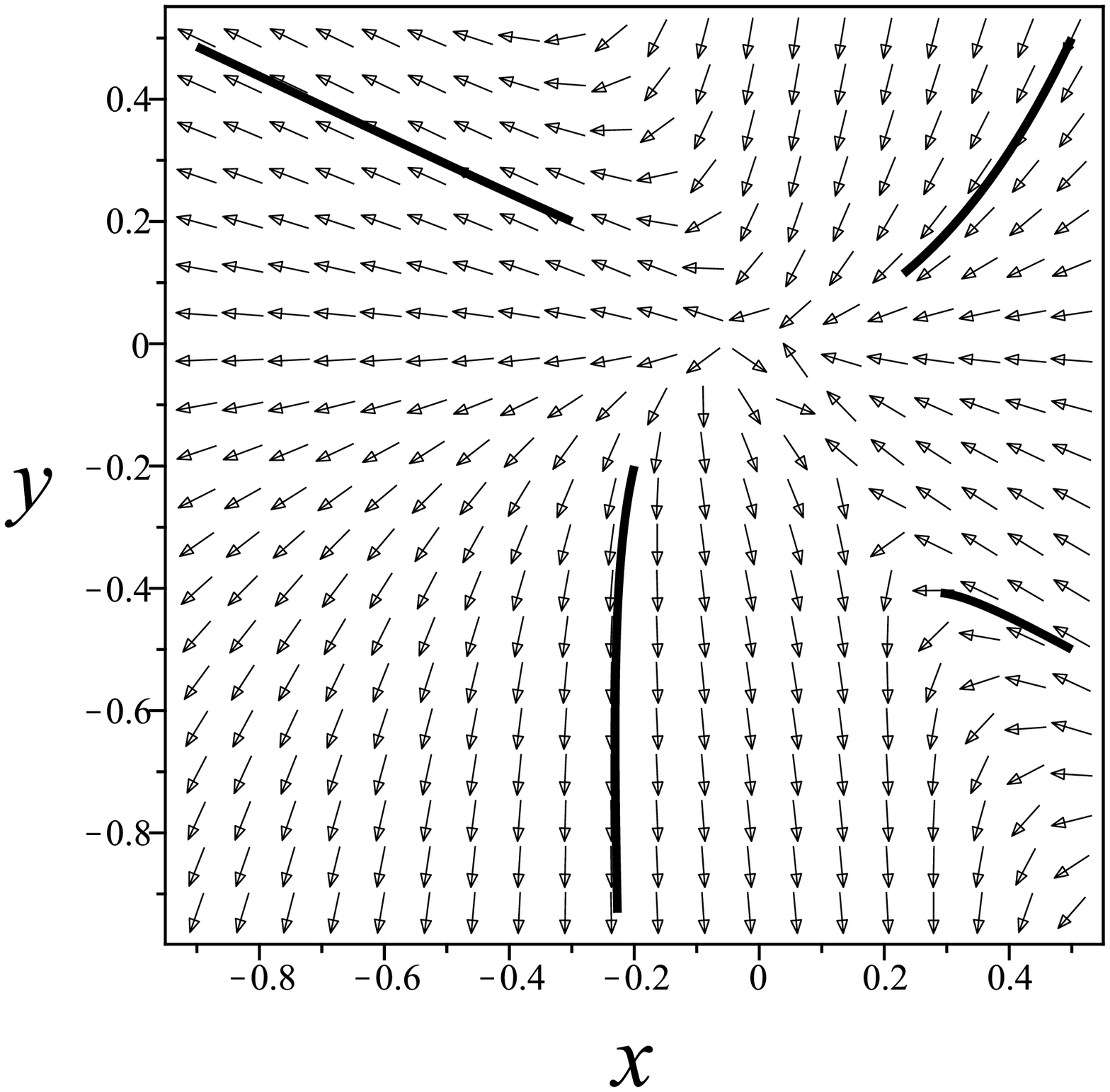}\hspace{1.5cm}
\includegraphics[width = 2.25in]{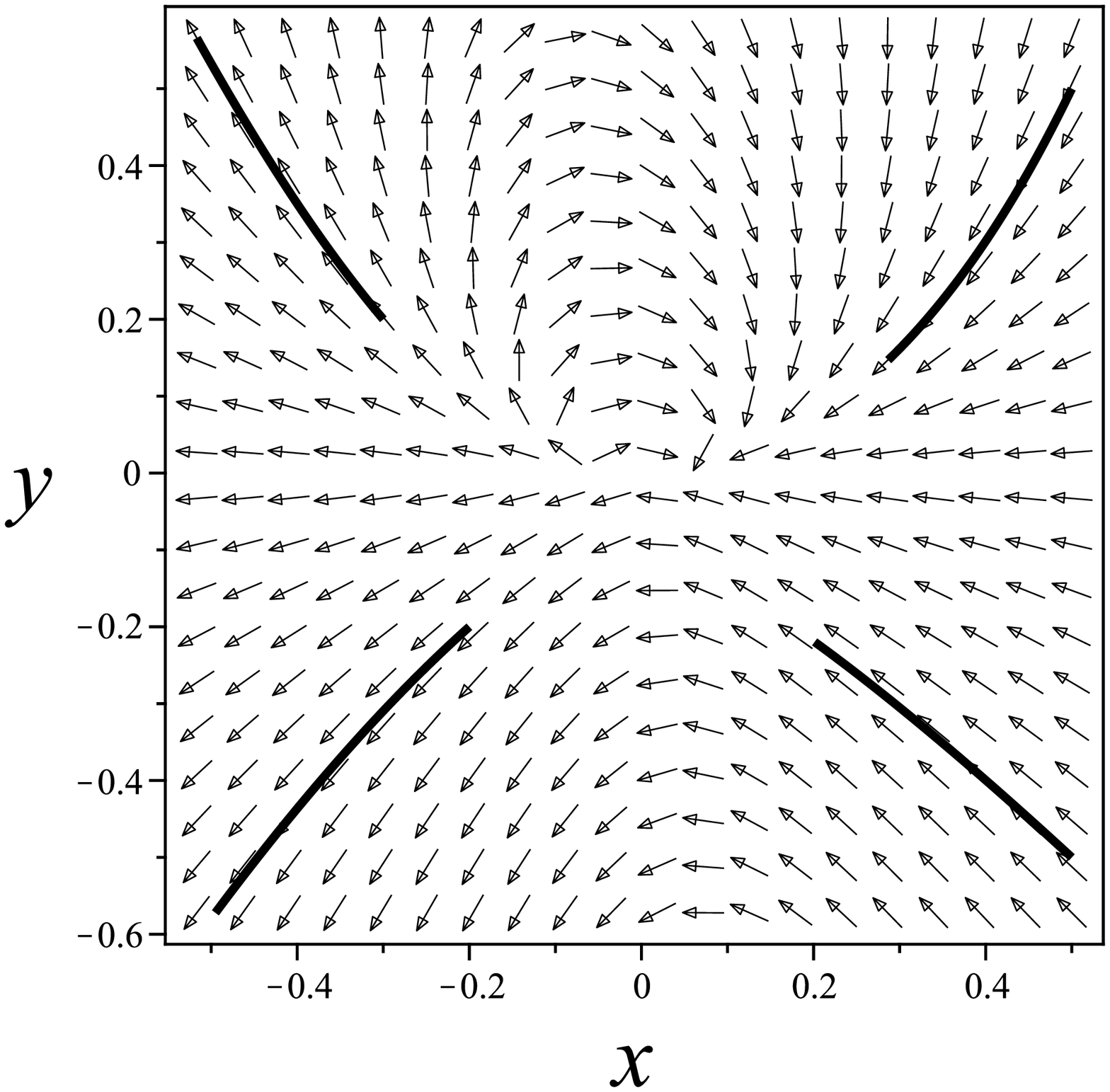}
\caption{$x-y$ solution spaces at $\omega = 1$ with $m=m_+$ (left plot) and $m=m_-$ (right plot). Four trajectories in both the figures are drawn for same 
initial conditions, which are -- $\{x(0) = 0.5, y(0) = 0.5\}$, $\{x(0) = 0.5, y(0) =- 0.5\}$, $\{x(0) = -0.3, y(0) = 0.2\}$ and $\{x(0) = -0.2, y(0) = -0.2\}$, while $t$ runs from 0 to 1.} \label{fig:phase}
\end{figure}

It is evident that Eq. \ref{eq:scalarBDtime} and Eq. \ref{eq:traceBDtime} 
constitute a dynamical system of their own for every possible value 
of $\omega$. At the critical point (0,0), the Jacobian of the linearised system vanishes. Therefore, let us directly look at the phase portraits 
(Fig. \ref{fig:phase}) of this system for $m=m_+$ (left plot) and 
$m=m_-$ (right plot) at, for example, $\omega = 1$.
Cosmologically viable features are apparent in the x-positive (expanding brane) and y-negative (decaying extra dimension) quadrant of both the phase spaces. 
In the left-side plot, one class of trajectories flow towards the centre, 
as time increases. Another class of trajectories flow toward $y = -\infty$ 
(extra dimension decaying at an ever increasing rate) solution as 
$t \rightarrow \infty$, while the $x$-coordinate increases very less. 
Now, $x$ becoming almost constant, in turn implies an almost exponential 
growth of $a(t)$ or brane inflation. In this case, this can mean that the 
brane expands very rapidly and forever -- a big rip.  

On the other hand, in the right-side plot, one class of x-positive and 
y-negative solutions flow towards the centre , which represents a static 
universe and another class of  trajectories flow toward $x = -\infty$ and 
$y = -\infty$ (whole of the space is singular at this point) while crossing 
the y-axis ($x = 0$) very slowly. This means the expansion of the brane 
first slows down to staticity and then starts contracting -- a big crunch. 
Another crossover takes place from x-negative and y-positive quadrant 
to x-positive and y-positive quadrant, for certain initial conditions, and 
then these trajectories flow towards the centre. Some part of these 
trajectories become parallel to $y$-axis, i.e. $x$ become constant, for a very 
short span of time, which can cause brane inflation, though the extra dimension
is also of the growing type.

\subsubsection{The vacuum solution}

Now, let us assume
\begin{equation}
y = h x ,\label{eq:linearity}
\end{equation}
i.e. we are considering the straight lines in the phase spaces at all values of $\omega$. Then the consistency requirement between Eq. \ref{eq:scalarBDtime} and Eq. \ref{eq:traceBDtime} leads to the following condition
\begin{equation}
A h^2 + (m + 1) h^2 + 1 = 0 \hspace{0.5cm} \Rightarrow \hspace{0.5cm} h_{\pm} = \frac{-(m+1) \pm \sqrt{(m+1)^2 - 4 A}}{2 A}. \label{eq:hcon}
\end{equation}
When the abovementioned constraint is satisfied, our job reduces to solving
only one independent equation
\begin{equation}
\dot x  = (A h^2 - 2) x.\label{eq:ind_x}
\end{equation}
Integrating further, we eventualliy get
\begin{equation}
a(t) \sim t^{\frac{1}{2 - A h^2}} \hspace{1cm} \mbox{and} \hspace{1cm} \eta(t) \sim t^{\frac{h}{2 - A h^2}}.\label{eq:vacsol}
\end{equation}
The most interesting feature of this specific set of solutions is that, 
these are actually {\em vacuum solutions}, i.e. one can easily check that 
all the matter energy momentum tensor components identically vanish when 
Eq. \ref{eq:hcon} is satisfied. The following figure gives a view of the 
exponents in $a(t)$ ($\nu_1 = \frac{1}{2 - A h^2}$) and $\eta(t)$ ($\nu_2 = \frac{h}{2 - A h^2}$) as functions of $\omega$.
\begin{figure}[!ht]
\includegraphics[width = 2.75in]{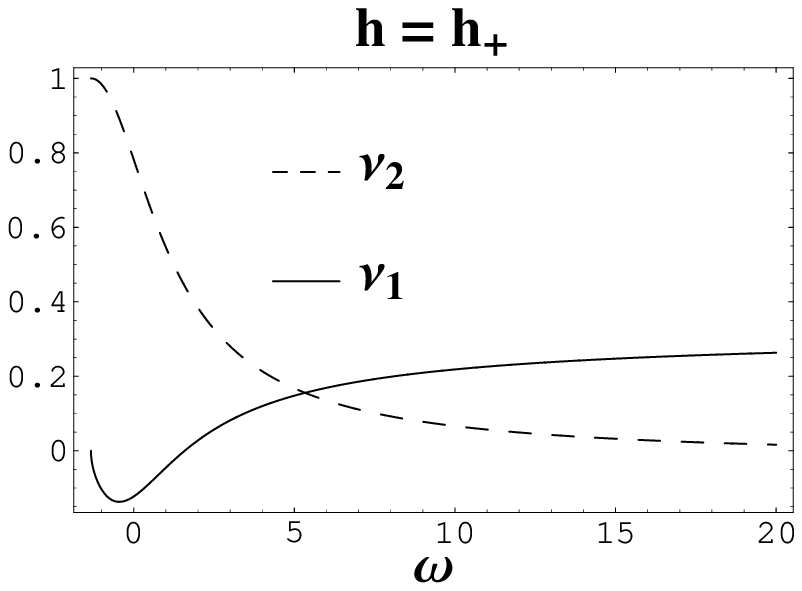}\hspace{1cm}
\includegraphics[width = 2.75in]{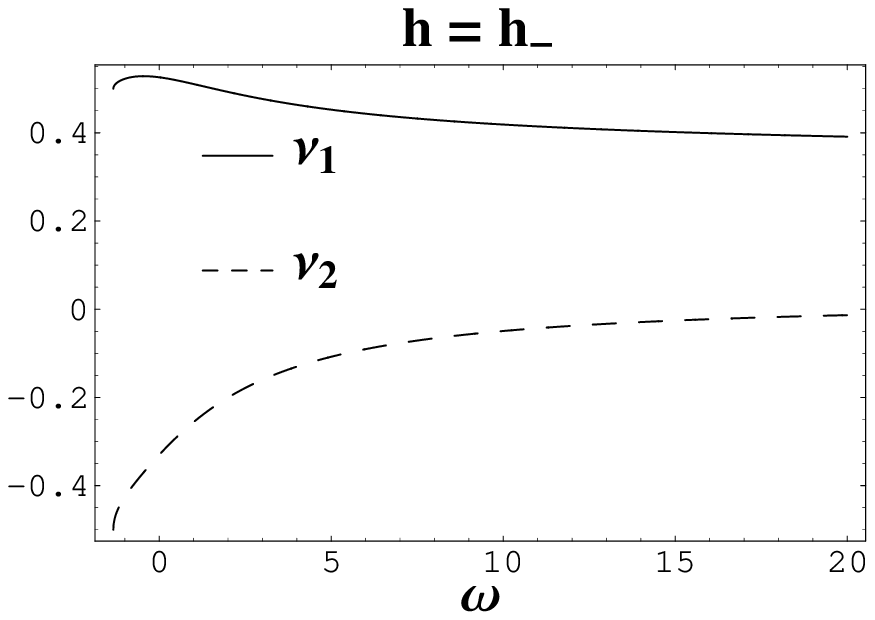}
\caption{Variation of the exponents, $\nu_1$ and $\nu_2$, of the scale factors 
$a(t)$ and $\eta(t)$ with respect to $\omega$ for $h = h_+$ and $h = h_-$ 
are shown in the left and right-side plots respectively.} \label{fig:BDvac}
\end{figure}
It can be clearly seen that all possible combinations of $\nu_1$ and $\nu_2$ (+ve+ve, +ve-ve, -ve+ve and -ve-ve) are there at different values of $\omega$. 
In fact, the variations seem to suggest that  
an expanding brane mostly suits a decaying extra dimension. 

\subsubsection{The radiative solution}

\begin{figure}[!ht]
\includegraphics[width = 2.5in]{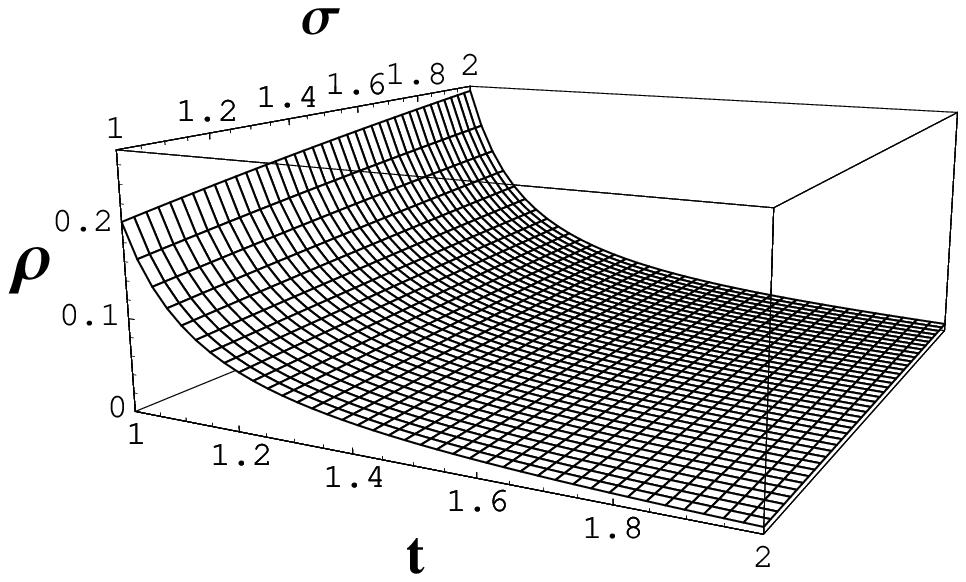}
\includegraphics[width = 2.5in]{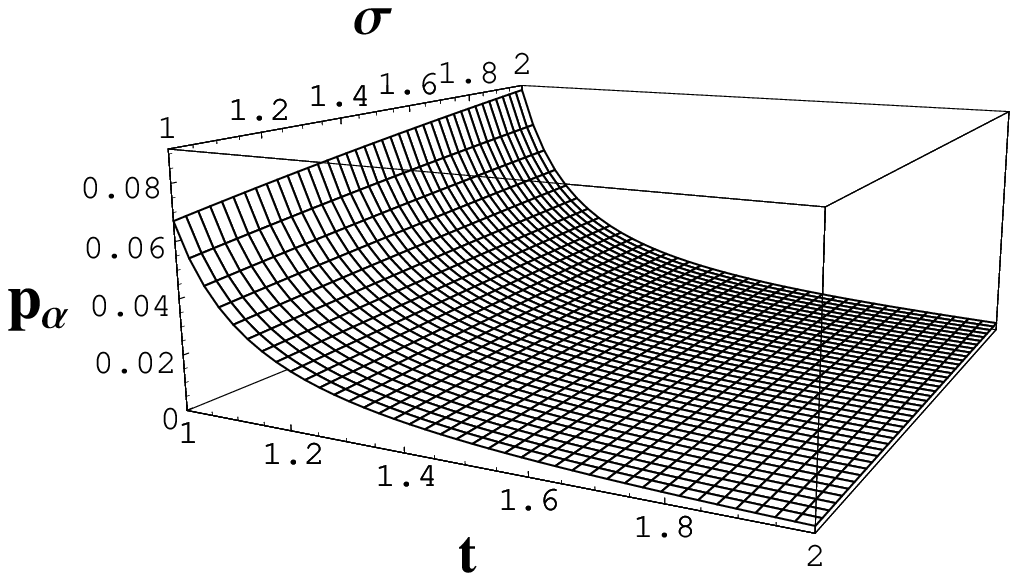}
\caption{Behaviour of the nonzero components of energy momentum tensor for bulk matter with radiative brane (using $m = m_+$ and $n = n_+$) with $c_1 = 0$ and $c_2 = 1$.} \label{fig:matter}
\end{figure}

One specific case, where an analytic solution is possible, is when $A = 0$, i.e. for $\omega = 1.586$, we have
\begin{eqnarray}
\dot x &=& - 2 x^2 \\
\dot y &=& - 3 x y - 2.26 y^2  .\label{eq:radiative}
\end{eqnarray}
Which implies,
\begin{eqnarray}
x &=& \frac{1}{2t - c_1} \label{eq:solradiative_x}\\
\mbox{and} \hspace{1cm} y &=& \frac{1}{(2t - c_1)(c_2 \sqrt{2t - c_1} - 2.26)}  ,\label{eq:solradiative_y}
\end{eqnarray}
where $c_1$ and $c_2$ are integration constants, whose different values will 
span the whole solution space of $x$ and $y$ for the abovementioned value of 
$\omega$. It may also be noted that, with $c_2 = 0$, we have the vacuum 
solution for $\omega = 1.586$ (in which case the extra dimension is essentially decaying again). Further, integrating Eq. \ref{eq:solradiative_x} and Eq. \ref{eq:solradiative_y} we get,
\begin{eqnarray}
a(t) &\sim& (2t - c_1)^{\frac{1}{2}} \label{eq:solradiative_a}\\
\mbox{and} \hspace{1cm} \eta(t) &\sim& \frac{(2.26 - c_2\sqrt{2t - c_1})^{0.44}}{(2t - c_1)^{0.22}}. \label{eq:solradiative_eta}
\end{eqnarray}
The above solution, in fact, represents a {\em radiative} brane while the 
nature of the extra dimension depends on $c_1$ and $c_2$. Fig. \ref{fig:matter} shows how the nonzero components of matter energy momentum tensor, $\rho$ 
and $p_{\alpha}$ ($p_{\sigma}$ vanishes in this case), behave as function 
$t$ and $\sigma$ with $c_1 = 0$ and $c_2 = 1$.

\subsubsection{Solutions in Brans-Dicke frame using Einstein frame scalar field solutions}

It is well known that the action of Brans--Dicke theory, as stated in the
previous section, can be converted into that of canonical Einstein gravity
coupled to a massless scalar field. We recall below how this is done.
First, let us rewrite the action for Brans Dicke theory replacing $\phi$
by $e^{-2{\tilde{\phi}}}$ and $g_{IJ}$ by ${\tilde{g}}_{IJ}$ . Subsequently, defining a conformally related metric
\begin{equation}
{{\tilde{g}}}_{IJ}= \Omega^2 (\sigma,t) g_{IJ}
\end{equation}
and choosing $\Omega = e^{\frac{2}{3}{\tilde \phi}}$ we find that the
Brans--Dicke action (with no additional matter fields) goes over to:
\begin{equation}
S=\int d^5 x \sqrt{-g} \left [ R- \frac{4}{3} \left (4+3\omega\right )g^{IJ}
\nabla_I {\tilde \phi} \nabla_J {\tilde \phi} \right ].
\end{equation}
Further, defining:
\begin{equation}
\Phi = \frac{2}{3}\sqrt{4+3\omega} {\tilde \phi}
\end{equation}
we can convert the action above into that of Einstein gravity coupled
to an ordinary scalar field.

This equivalence can now be used to construct new solutions in Brans--Dicke theory
by making use of the solutions with an ordinary scalar field discussed earlier in Section {\bf IV B}.
The main point here is that the solutions with an ordinary scalar are also
solutions of Brans--Dicke theory in the Einstein canonical frame. How do
these solutions look like in the Brans--Dicke frame? We look at this
aspect now.

Following the earlier ordinary scalar field analysis we choose:
\begin{equation}
\Phi = m \ln \eta +\frac{3}{m} f
\end{equation}
with $a(t)\sim (t-t_0)^{\frac{1}{n+3}}$, $\eta(t) \sim (t-t_0)^{\frac{n}{n+3}}$
and $f(\sigma)= \frac{1}{4} \ln (\sigma-\sigma_0)$.
We also had $m^2=\frac{3}{4}$ and $n=4\pm 2\sqrt{6}$.

\begin{figure}[!ht]
\includegraphics[width = 5in]{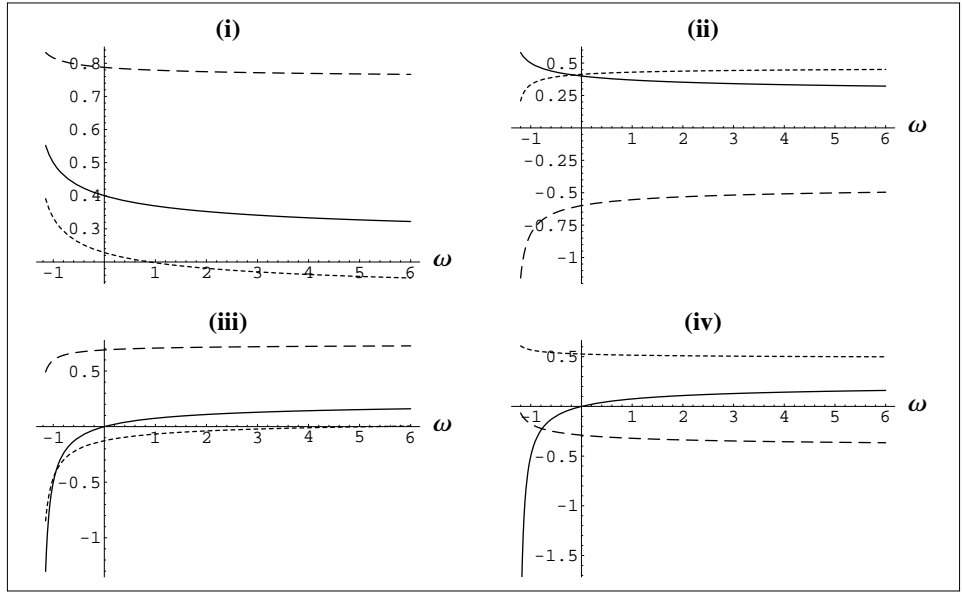}
\caption{Variations of the coefficient $\left[\frac{4q+1}{4q+4}\right]$ in $f_1 (\sigma_1)$ (continuous line), the exponent $\left[\frac{p(n+3) + 1}{(p+1)(n+3)}\right]$ in the cosmological scale factor $a_1(\tau)$ (dotted lines) and the exponent $\left[\frac{p(n+3) + n}{(p+1)(n+3)}\right]$ in the extra dimensional scale factor $\eta_1(\tau)$  (dashed lines) w.r.t. $\omega$ are shown with following different (m, n) combinations; set{\bf(i)}: $\left(\frac{\sqrt{3}}{2}, 4 + 2 \sqrt{6}\right)$, set{\bf(ii)}: $\left(\frac{\sqrt{3}}{2}, 4 - 2 \sqrt{6}\right)$,  set{\bf(iii)}: $\left(\frac{-\sqrt{3}}{2}, 4 + 2 \sqrt{6}\right)$ and set{\bf(iv)}: $\left(\frac{-\sqrt{3}}{2}, 4 - 2 \sqrt{6}\right)$.}\label{fig:BD_omega} 
\end{figure}

The metric in the Brans--Dicke frame is related to that in the Einstein
frame by an overall conformal factor $\Omega^2= e^{\frac{4}{3}{\tilde{\phi}}}
$. Using the relation between $\tilde{\phi}$ and $\Phi$ we find that the
conformally related line element becomes:
\begin{equation}
ds^2 = e^{2f_1(\sigma_1)} \left [ -d\tau^2 +a_1^2(\tau){\vert d\vec x\vert}^2
\right ] + \eta_1^2 (\tau) d\sigma_1^2
\end{equation}
where:
\begin{eqnarray}
\tau - \tau_0 =\frac{(t-t_0)^{p+1}}{p+1} \hspace{0.1in};\hspace{0.1in} \sigma_1 - \bar \sigma_1= \frac{(\sigma-\sigma_0)^{q+1}}{q+1}\\
p= \frac{mn}{(n+3)\sqrt{3(4+3\omega)}} \hspace{0.1in};\hspace{0.1in}
q= \frac{3}{4m\sqrt{3(4+3\omega)}}
\end{eqnarray}
and
\begin{eqnarray}
f_1 (\sigma_1) &=& const. + \frac{4q+1}{4q+4} \log(\sigma_1 - \bar \sigma_1), \label{eq:f_1}\\
a_1(\tau) &\sim& (\tau - \tau_0)^{\frac{p(n+3) + 1}{(p+1)(n+3)}},\label{eq:a_1}\\
\eta_1(\tau) &\sim& (\tau - \tau_0)^{\frac{p(n+3) + n}{(p+1)(n+3)}}.\label{eq:eta_1}
\end{eqnarray}
We now illustrate the nature of the coefficient in  $f_1$ and the exponents in $a_1$ and $\eta_1$ appearing in the above solutions as functions of $\omega$ for the four possible combinations of (m,n) through Fig. \ref{fig:BD_omega}.
 It is interesting to note that for set{\bf(i)} and  set{\bf(iii)} growing $a_1(\tau)$ - decaying $\eta_1(\tau)$ combination does not exist whereas for  set {\bf(ii)} and  set{\bf(iv)} both decaying and growing warp factor solutions exist with desired evolutions for $a_1(\tau)$ and $\eta_1(\tau)$.
\subsection{Bulk dilaton scalar}

Low energy effective string theory gives rise to Einstein-like equations
through the conditions that the $\beta$-functions of the string $\sigma$-model
are equal to zero \cite{string}. The Einstein like equations have 
additional terms
involving the dilaton (a scalar), third rank antisymmetric tensor 
and other fields (Maxwell and moduli fields) which arise out of
the method of compactification. Dilaton gravity involves only the
dilaton and the metric field. It is different from ordinary scalar field
theory and is equivalent to Brans-Dicke theory under a special
choice of the  $\omega$ parameter (being set equal to -1). The action for dilaton gravity \cite{dilaton} is given as:
\begin{equation}
S = \int{d^5x \sqrt{-g} e^{-2\phi} (R + 4 \partial_K\phi \partial^K\phi)}. \label{eq:actiondilaton}
\end{equation}
This gives the following field equations,
\begin{eqnarray}
G_{IJ} = - 2 \nabla_I\nabla_J\phi + g_{IJ} \nabla^2\phi, \label{eq:einsteindilaton}\\
g^{IJ} \nabla_I \nabla_J \phi = 2 g^{IJ} \nabla_I \phi \nabla_J \phi.
\end{eqnarray}

\subsubsection{A solution in the string frame}

The terms in the R. H. S. of the above equation can be clubbed together to
yield an effective energy momentum tensor. The effective matter stress energy
is therefore given as:
\begin{eqnarray}
\mbox{ie.} \hspace{1cm} T_{00} &=& e^{-2f} (- \ddot\phi + (3x + y)\dot\phi) - \frac{1}{\eta^2}(\phi'' + 2f'\phi'), \nonumber \\
T_{\alpha\alpha}  &=& - e^{-2f} (\ddot\phi + (x + y)\dot\phi) + \frac{1}{\eta^2}(\phi'' + 2f'\phi'), \nonumber \\
T_{44} &=& - e^{-2f} (\ddot\phi + (3x - y)\dot\phi) + \frac{1}{\eta^2}(- \phi'' + 4f'\phi')\nonumber \\
\mbox{and} \hspace{1cm} T_{04} &=& \frac{e^{-f}}{\eta}(- 2\partial_t\partial_\sigma\phi + 2f'\dot\phi + 2y\phi'). \label{eq:TijDilaton} 
\end{eqnarray}
Let us assume, as before,
\begin{equation}
\phi(t,\sigma) \equiv \phi_1(t) + \phi_2(\sigma). \label{eq:phiform} 
\end{equation}
Then the off--diagonal term in the Einstein equations gives us the
following constraints, 
\begin{equation}
\dot\phi_1 = m y \hspace{.5cm}\mbox{and}\hspace{.5cm} \phi_2' = n f' \hspace{.5cm}\mbox{where}\hspace{.5cm} m = \frac{3 - 2n}{2}. \label{eq:phiconstraintD}
\end{equation}
So, essentially we have,
\begin{equation}
\phi(t,\sigma) = m\log[\eta(t)] + n f(\sigma) + const.  \label{eq:phisolD} 
\end{equation}

Now, equating the coefficients of $\frac{1}{\eta^2}$ in the both sides of the Einstein's equations we get,
\begin{eqnarray}
f''(\sigma) + \frac{6 - 4n}{n}f'^2(\sigma) = 0 \\
\hspace{.5cm}\mbox{and}\hspace{.5cm} (m,n) \equiv \left(\frac{1}{2},1\right) \hspace{.3cm}\mbox{or}\hspace{.3cm} \left(-\frac{3}{2},3\right).
\end{eqnarray}
On the other hand, equating the coefficient of $\frac{1}{\eta^2}$ in the
scalar field equation gives:
\begin{equation}
f''(\sigma) + \frac{4- 2n}{n}f'^2(\sigma) = 0 
\end{equation}
Thus the only allowed value for (m,n) is $m=\frac{1}{2}$ and $n=1$ which
gives:
\begin{equation}
f(\sigma) = const. + \frac{1}{2}\log(\sigma - \sigma_0), \hspace{1cm}\mbox{a growing warp factor,} 
\end{equation}

Finally, equating the the coefficients of $e^{-2f}$ in the Einstein
equations (with $m=\frac{1}{2}$), we get,
\begin{eqnarray}
3 x^2 + \frac{3}{2} x y  + \frac{1}{2} \dot y - \frac{1}{2} y^2 &=& 0, \nonumber  \\
2\dot x + 3x^2 + \frac{3}{2} x y + \frac{1}{2}\dot y + \frac{1}{2} y^2 &=& 0  \nonumber \\ 
\mbox{and}\hspace{1cm} 3\dot x + 6x^2 - \frac{1}{2}\dot y - \frac{3}{2} 
x y + \frac{1}{2} y^2 &=& 0.
\end{eqnarray}
The three coupled equations above give rise to the following algebraic constraint,
\begin{equation}
3 x^2 - \frac{1}{2} y^2 = 0 .\label{eq:extraconstraintD} 
\end{equation}
which yields
\begin{equation}
y = \pm\sqrt{6} x, \hspace{.5cm} \frac{\ddot a}{a} + 2 \frac{\dot a^2}{a^2} = 0 \hspace{.5cm}\mbox{and}\hspace{.5cm} \frac{\ddot \eta}{\eta} + \left(\pm\sqrt{\frac{3}{2}} - 1\right) \frac{\dot \eta^2}{\eta^2} = 0, \label{eq:1steqnsD}
\end{equation}
with solutions as,
\begin{equation}
 a(t) \sim (t - t_0)^{\frac{1}{3}} \hspace{.5cm}\mbox{and}\hspace{.5cm} \eta(t) \sim (t - t_0)^{\pm\sqrt{\frac{2}{3}}}.\label{eq:1stsolns}
\end{equation}
Note that the coefficient of $e^{-2f}$ in the scalar field equation gives
$\dot y + 3 x y=0$, which is automatically satisfied by the abovementioned
solution. We discard the positive exponent solution for $\eta$, on physical grounds discussed before.

\subsubsection{Solutions in string frame using Einstein-scalar solutions in
Einstein frame}

It has been noted earlier that for $\omega=-1$, Brans--Dicke theory
gives rise to dilaton gravity. Using the solutions for Einstein
gravity coupled to an ordinary scalar we now construct string (Brans--Dicke)
frame solutions in dilaton gravity, following the discussion presented in {\bf IV B}.

Substituting $\omega=-1$ in Eqns. \ref{eq:f_1}-\ref{eq:eta_1} we get the following forms for $a_1(\tau)$,
$\eta_1(\tau)$ and $f_1(\sigma_1)$.
\begin{eqnarray}
f_1 (\sigma_1) = const. + \frac{4q+1}{4q+4} \log(\sigma_1 - \bar \sigma_1), \\
a_1(\tau) = \left[ (p+1)(\tau - \tau_0\right ]^{\frac{p(n+3) + 1}{(p+1)(n+3)}},\\
\eta_1 (\tau) =\left[ (p+1)(\tau - \tau_0)\right ]^{\frac{p(n+3) + n}{(p+1)(n+3)}}.
\end{eqnarray}

with $p=\frac{mn}{\sqrt{3} (n+3)}$, $q=\frac{\sqrt{3}}{4m}$ and $m=\pm \frac{\sqrt{3}}{2}$,
$n=4\pm 2 \sqrt{6}$.

\begin{figure}[!ht]
\includegraphics[width = 5in]{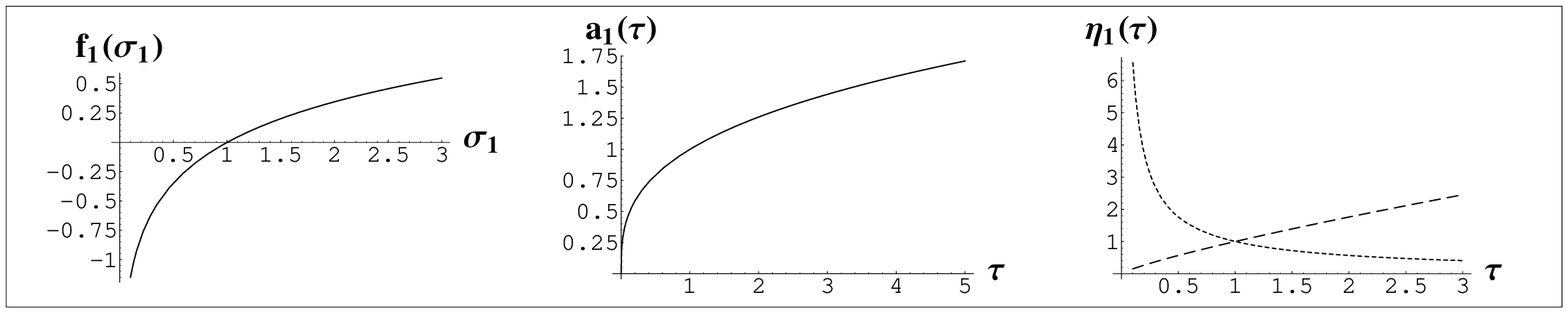}
\includegraphics[width = 5in]{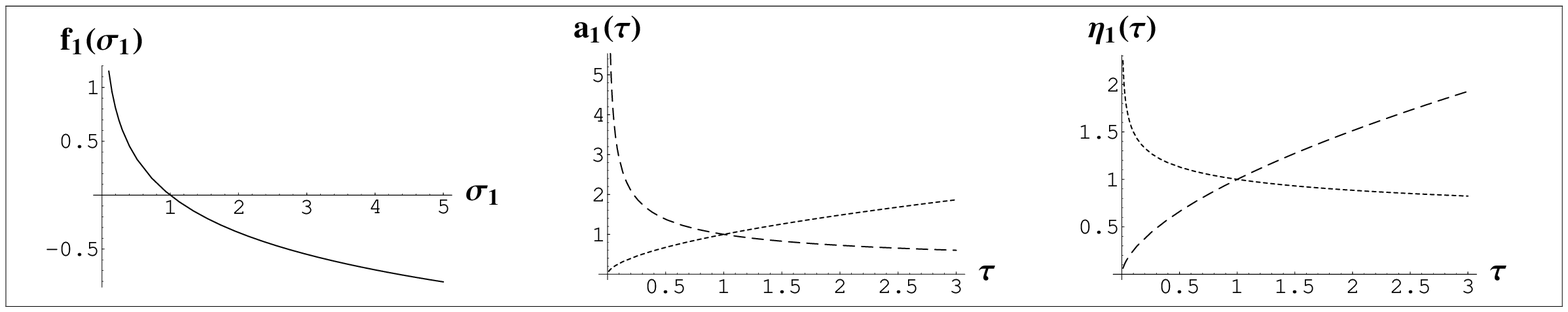}
\caption{Nature of the possible combinations of $a_1(\tau)$, $\eta_1(\tau)$ and $f_1(\sigma_1)$ found in presence of a bulk dilaton field. In the top row: $(m,n) = (\frac{\sqrt{3}}{2},4\pm 2 \sqrt{6})$ and in the bottom row: $(m,n) = (\frac{- \sqrt{3}}{2},4\pm 2 \sqrt{6})$. The exponents in $a_1(\tau)$ are same for  $(m,n) = (\frac{\sqrt{3}}{2},4 + 2 \sqrt{6})$ and $(m,n) = (\frac{\sqrt{3}}{2},4 - 2 \sqrt{6})$.}\label{fig:Dplots} 
\end{figure}

The four sets for $a_1$, $\eta_1$ and $f_1$ (correspomding to four different combination of ($m,n$)) yield two acceptable solutions -- one with a decaying warp factor and another with a growing warp factor (this is in fact the same solution as given by Eq. \ref{eq:1stsolns}).
Fig.\ref{fig:Dplots} shows the nature of these solutions in presence of bulk dilaton field.

The explicit solutions (those which are physically meaningful) are displayed in Table I.

\section{Status of energy conditions for the above solutions}
We now focus our attention on analysing the energy conditions some of the above solutions.

\begin{figure}[!ht]
\includegraphics[width = 5 in]{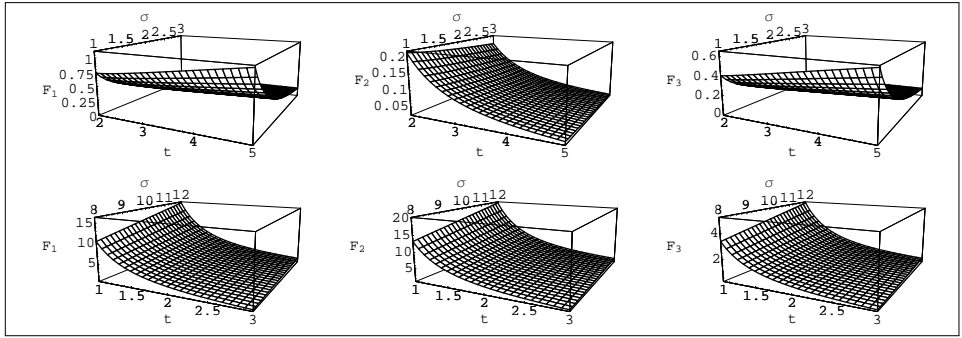}
\caption{Status of the inequalities for the two ``good'' solutions 
as given in the previous table for $t_0 = 0$ and $\sigma_0 = 0$ with 
suitably chosen $t_c$ and $\sigma_c$.} \label{fig:SolPlots} 
\end{figure}

\begin{table}[!ht]
\begin{center}
\begin{tabular}{|c|c|c|c|c|c|c|}
\hline
Bulk Field & $f(\sigma) = const. +  $         & $a(t)\sim $      & $\eta(t)\sim $    & $F_1\ge 0$ & $F_2 \ge 0$ & $F_3 \ge 0$\\
\hline 
Ordinary Scalar & $0.25\log(\sigma -\sigma_0)$  & $(t-t_0)^{0.48}$ & $(t-t_0)^{-0.43}$  &$\surd$ &$\times $& $\surd$ \\\cline{3-7}

\hline 

\multirow{2}*{Brans-Dicke (for $\omega = 1$)}

&$0.37\log(\sigma -\sigma_0)$    & $(t-t_0)^{0.43}$  & $(t-t_0)^{-0.55}$  &$\surd$  &$ \times$  &  $\surd$     \\\cline{2-7}

&$0.075\log(\sigma - \sigma_0)$ & $(t-t_0)^{0.515}$ & $(t-t_0)^{-0.32}$   & $\surd$  & $\surd$ & $\surd$      \\
                                                        
\hline 
\multirow{2}*{Dilaton Scalar ($\omega = -1$)}
           & $0.5\log(\sigma -\sigma_0)$      & $(t-t_0)^{0.33}$ & $(t-t_0)^{-0.82}$    &$\times$  & $\times$ &$\surd$ \\\cline{2-7}
                                                                                         
           &$-0.5\log(\sigma-\sigma_0)$     & $(t-t_0)^{0.57}$ & $(t-t_0)^{-0.18}$  &  $\surd$&$\surd$  & $\surd$      \\

\hline 
\end{tabular}
\caption{The above table shows few of the typical solutions, found in Section 
IV, mentioning their desirability as viable models and the status of the energy condition inequalities (Eqs. \ref{eq:ineq1}, \ref{eq:ineq2}, \ref{eq:ineq3}).}\label{tab:solutions}
\end{center}
\end{table}
In Table \ref{tab:solutions}, we have listed the `good' solutions (only those 
which have a combination of growing cosmological scale and a 
decaying extra dimensional scale) found in the previous sections.
For the Brans-Dicke case
,
we have two solutions for $\omega =  1$. The solutions with a dilaton field are essentially derived from general solutions for the Brans-Dicke case by just 
equating $\omega = -1$. There are solutions with both decaying and growing warp factors. Notably, all solutions have singularities at a finite value
of $t$ or $\sigma$. To investigate the status of WEC for these solutions, 
we have plotted the inequality functions. It is found that the 3rd and 5th 
set of solutions satisfy all the inequalities (Fig.\ref{fig:SolPlots}). 
Here, the statement {\em inequality is satisfied } means that it is satisfied 
in an entire semi-infinite spacetime region from 
\{($t = t_c, \sigma = \sigma_c$) to ($t = \infty, \sigma = \infty)$\}, where $t_c$ and $\sigma_c$ are suitably chosen lower bounds (these bounds essentially comes from the constraint inequality relation \ref{eq:bounds}) on the allowed 
domain for $t$ and $\sigma$ (energy condition satisfying solutions have to be 
defined in these domains only). The statement {\em not satisfied} means 
that no semi-infinite spacetime region can be found where energy conditions are satisfied.

\section{Placing the Brane and Singularity Resolution}

Till now we have essentially found bulk solutions. To have a proper 
brane-bulk system, we need to place a four dimensional hypersurface in the bulk. One can do this by replacing the argument of the {\em logarithm} (appearing in every solution of $f(\sigma)$), i.e. $\sigma$, by $|\sigma|$. The contribution of this {\em modulus} function will appear as a Dirac 
delta function peaked at some $\sigma_b$ (which is the location of the
brane), in the left hand side of Einstein 
equations. To justify this, the energy momentum tensor on the right hand side of Einstein 
equations must carry a delta-peaked term too. This can be done by adding an extra term multiplied with a delta function in the bulk Lagrangian, which also peaks at the location where we intend to place the thin brane.

\subsection{Junction conditions}

Let us take the contribution of the brane to the Einstein-Hilbert action to be,
\begin{equation}
S_b = -\int{d^4x \sqrt{-g_4}\ \lambda(\phi)} \equiv -\int{d^5x \sqrt{-g_5}\ \lambda(\phi)} \frac{\delta(\sigma - \sigma_b)}{\sqrt{g_{44}}}. \label{eq:branetension}
\end{equation} 
Here $g_5$ denotes the determinant of the full five-dimensional bulk metric and $g_4$ is the determinant of the induced metric on the brane. Now, we can 
determine the jump in $f'(\sigma)$ (or $\phi'(\sigma)$) across the brane, 
from the junction conditions \cite{israel}, in the
model cases of the ordinary scalar field or the Brans-Dicke field coupled to gravity.

\subsubsection{Junction conditions with ordinary scalar field}
Integrating the scalar field equation at the location of the brane, we get,
\begin{equation}
\big[\phi_2'\big] = \frac{\eta}{2}\frac{d\lambda}{d\phi}\ \Bigg\vert_{\sigma = \sigma_b} \label{eq:scalar_phi1}
\end{equation}
Then from Einstein equations we have,
\begin{equation}
\big[f'\big] = - \frac{\eta}{6}\lambda\ \Bigg\vert_{\sigma = \sigma_b} \label{eq:scalar_f}
\end{equation}
Now, using the constraint $\phi_2' = \frac{3}{m} f'$, we obtain,
\begin{equation}
\lambda(\phi) = \lambda_0\ e^{-\frac{\phi}{m}}, \label{eq:scalar_lambda}
\end{equation}
where $\lambda_0$ is an integration constant. For the above expression of $\lambda(\phi)$, junction conditions \ref{eq:scalar_phi1} and \ref{eq:scalar_f}, in fact, become independent of $\eta(t)$ (similar feature is reported in \cite{koyama}).

\subsubsection{Junction conditions with Brans-Dicke field} 
Similarly, in the Brans-Dicke case, the scalar field equation (Eq. \ref{eq:scalareqnbd1}) at the location of the brane  (with the brane tension given by Eq. \ref{eq:branetension}) gives us,
\begin{equation}
-4 \big[f'\big] + 2\omega \frac{\big[\phi_2'\big]}{\bar \phi} = \frac{\eta}{2}\frac{d\lambda}{d\phi}\ \Bigg\vert_{\sigma = \sigma_b}. \label{eq:BD_junc1}
\end{equation}
where $\bar\phi$ is, the mean value of the function $\phi(t,\sigma)$ at $\sigma = \sigma_b$, defined as
\begin{equation}
\bar\phi = \frac{\phi(\sigma_b^+) + \phi(\sigma_b^-)}{2}. \label{eq:BD_phiav}
\end{equation}
Further, from the $00$ or $\alpha\alpha$ component of the 
Einstein equations, we get,
\begin{equation}
3 \big[f'\big] = - \frac{\eta}{2}\lambda\ \Bigg\vert_{\sigma = \sigma_b}, \label{eq:BD_junc2}
\end{equation}
Using the relation $\frac{\phi'}{\phi} = n f'$, leads us to
\begin{equation}
\lambda(\phi) = \lambda_0\ e^{-\frac{2n\omega - 4}{3}\phi}, \label{eq:BD_lambda}
\end{equation}
for which the junction condition becomes,
\begin{equation}
\big[f'\big] = - \frac{\lambda_0}{6} \eta\ e^{-\frac{2n\omega - 4}{3}\bar\phi}.  \label{eq:BD_juncf}
\end{equation}
In this case, the junction conditions are time dependent.

The junction conditions for the dilaton case can be obtained by
using $\omega=-1$ and a redefinition of the $\phi$ field in the
above analysis for the Brans--Dicke scalar.

\subsection{Resolving bulk singularities}

It may be noted that all our solutions (except those discussed under the
section on an exponential RS type warp factor) have bulk singularities, apart from the
usual big-bang cosmological singularity in time. The latter is inevitable
and expected while the former needs better understanding.

The main question we need to address is -- how do we manage to use a
spacetime with a bulk singularity as a model for our bulk five
dimensional geometry? There are ways to do this.

One possible resolution of the singularities is done by putting branes 
at the locations of the singularities. For example, let us take, in the case of 
solutions with a bulk dilaton field,  the two different solutions valid in 
two different regions (except the points where singularities occur), i. e.
\begin{equation}
f(\sigma) = \left\{\begin{array}{ll} \frac{1}{2}\log|\sigma - \sigma_1| + c_1, \hspace{1cm}\mbox{for} \hspace{1cm} \sigma \le \sigma_0
                                  \\ -\frac{1}{2}\log|\sigma - \sigma_2| + c_2, \hspace{1cm}\mbox{for} \hspace{1cm} \sigma \ge \sigma_0
                   \end{array} \right.
\end{equation} 
where $\sigma_0$ is located somewhere between $\sigma_1$ and $\sigma_2$. Then continuity at $\sigma_0$ implies,
\begin{equation}
\frac{1}{2}\log|\sigma_0 - \sigma_1| + c_1 = -\frac{1}{2}\log|\sigma_0 - \sigma_2| + c_2, 
\end{equation} 
Now this is a good bulk solution except at points $\sigma_1$ and $\sigma_2$. 
To resolve these singularities we can put two branes at $\sigma_1$ and 
$\sigma_2$ and another brane at $\sigma_0$ and make their total contribution 
to bulk vacuum energy vanish \cite{stefan_etal} (as we do not have any bulk cosmological constant). This will be  equivalent to putting delta-function source 
terms in the bulk Lagrangian. In effect, those singularities in bulk solutions actually provide us with places for the 3--branes, one of which can be chosen to be the Standard Model brane we live in.

\section{Conclusions}

Finally, we list below, systematically, the conclusions obtained in this investigation.
\begin{itemize}
\item We start out by writing down the energy condition inequalities using the Einstein 
tensor components (which have a nonzero $G_{04}$). These inequalities are first checked 
for typical choices for the various metric functions $a(t)$, $\eta(t)$ and $f(\sigma)$. We demonstrate that there exist viable
models (following the criteria listed in the Introduction) which satisfy
the energy conditions.
\item We also find exact solutions with matter sources of various kinds. 
To begin with, we look
at possible matter sources that may arise if we assume an exponential 
(in $\sigma$) warp factor and some
other typical constraints on matter stress energy. The solution space
for $[a(t),\eta(t)]$ is analysed using a dynamical systems approach. 
Subsequently, we look at solutions with various
types of scalar fields as sources -- eg. bulk normal scalar, the Brans-Dicke 
scalar and the dilaton scalar. In our approach to obtaining solutions we 
exploit the known fact that with a conformally related metric and a redefined 
scalar one can convert BD theory into canonical Einstein Gravity coupled to a 
scalar.
Analytic solutions are written down and the various possibilities that arise 
are outlined. In the case with a Brans-Dicke scalar and matter in the bulk, 
it is observed that, for traceless matter energy momentum 
exact and viable analytical solutions can be found with decaying or growing warp
factors as well as decelerating/accelerating $a(t)$. 
We note that in the case of the dilaton we can have decaying as well as growing 
warp factor solutions too. We are also able to find desirable solutions with 
growing $a(t)$ and decaying $\eta(t)$ in several of our examples.

\item We check the energy conditions, the nature of the functions $a(t)$ 
and $\eta(t)$ and figure out whether an obtained solution is desirable or not. 
The details are 
tabulated in one of the columns in the table commenting on the desirability of the solution. 
We do find several classes of desirable solutions, which satisfy our
requirements and the energy conditions.
\item Finally, we outline how one can place a brane in the above bulk
spacetimes by using the junction conditions . We also briefly discuss a
way to resolve the bulk singularities using standard
techniques.
\end{itemize}

With three different functions appearing in the line element, it is always very difficult to
find exact solutions. That we have found some is indeed encouraging. 
We hope to use these solutions in actual brane cosmological scenarios
and arrive at relevant conclusions by making use of
existing observational data on supernova and CMB anisotropies. We would also 
like to extend our results to other types of stress energy expressions. 
A disturbing aspect of our
solutions is that they are singular in the bulk, though we do provide ways of
resolving them. It remains to be seen whether we can find newer solutions
(a) which are not necessarily
of a power law type, (b) which have, in appropriate limits, both deceleration 
and acceleration, (c) which have a decaying
extra dimension stabilisable to a finite value and (d) where the 
bulk warp factor gives rise to a non-singular bulk metric, similar to RS. In essence, an appropriate combination of different bulk fields, with some dominating over the others in specific time intervals, will then be able to generate the expansion history (with proper decelerating and accelerating phase), of our universe on the brane.

\section*{Acknowledgements}
SG thanks IIT Kharagpur for providing financial support and Centre for
Theoretical Studies, IIT Kharagpur for its research facilities. We also thank P. S. Dutta for useful discussions.

\end{document}